\documentclass{iauc}
\usepackage{graphicx}

\title[Speckle nulling] 
{Speckle nulling \\ with space-based coronagraphs}

\author[Bord\'e \& Traub]   
{Pascal J. Bord\'e$^1$
  \thanks{Michelson Postdoctoral Fellow; present address: Michelson Science Center, California Institute of Technology, 770 S Wilson Avenue, MS 100-22, Pasadena, CA 91125.},
\and Wesley A. Traub$^2$}

\affiliation{$^1$Harvard-Smithsonian Center for Astrophysics, 60 Garden Street, Cambridge, MA 02138, USA\break 
email: pborde@cfa.harvard.edu \\[\affilskip]
$^2$Jet Propulsion Laboratory, M/S 301-451, 4800 Oak Grove Drive, Pasadena, CA 91109, USA \break 
email: wtraub@jpl.nasa.gov}

\pubyear{2006}
\volume{200} 
\pagerange{xxx--xxx}
\date{November 7th, 2005}
\jname{Direct Imaging of Exoplanets: Science \& Techniques}
\editors{C. Aime and F. Vakili, eds.}
\begin{document}

\maketitle

\begin{abstract}
Following the tracks of \cite{Malbet95} on dark hole algorithms, we present analytical methods to measure and correct the speckle noise behind an ideal coronagraph. We show that, in a low aberration regime, wavefront sensing can be accomplished with only three images, the next image being fully corrected (no iterative process needed). The only hardware required is the coronagraph deformable mirror and an imaging detector in the focal plane, thus there are no non-common path errors to correct. Our first method, \emph{speckle field nulling}, is a fast FFT-based algorithm requiring the deformable mirror influence functions to have identical shapes. Our second method, \emph{speckle energy minimization} is more general and based on matrix inversion. Numerical simulations show that these methods can improve the contrast by several orders of magnitude.
\end{abstract}

%
%
\firstsection
\section{Introduction} \label{sec:int}
By \emph{speckle nulling}, we refer to any technique whose goal is the reduction of speckle noise down to a level compatible with exoplanet detection. In the focal plane, speckle noise appears as a background made of a collection of star-like spots, the speckles. Speckles doubly hamper planet detection: they worsen the photon noise by increasing the background, and they introduce confusion because speckles and planets look alike.

Speckle noise arises from the scattering of star light by optical defects acting either as phase gratings (e.g. polishing errors), or as amplitude gratings (e.g. inhomogeneous reflectivity). Speckle nulling is achieved by creating destructive interferences at the locations of speckles thanks to a deformable mirror \cite[(DM for short; see Trauger, Burrows, Gordon, \etal\ 2004)]{Trauger04}. In terms of wave propagation, optical defects cause wavefront aberrations that the DM compensate for.

Speckle nulling belongs to the broader topic of wavefront sensing and control. As explained by \cite{Malbet95} in their pioneering work, it differs from conventional adaptive optics (AO) in the sense that its goal is not to maximize the Strehl ratio (i.e. minimize the standard deviation of phase errors), but to minimize the light scattered in a given zone of the image plane, dubbed \emph{dark hole}, where planet detection becomes possible. Moreover, conventional AO measures the wavefront with a separate sensing channel, most of the time at a wavelength shorter than the wavelength used for science operation. For high-contrast imaging, it is highly desirable to measure the wavefront directly with the science image in order to avoid non-common path errors.

In their paper, \cite{Malbet95} discuss only wavefront control assuming prior knowledge of the wavefront thanks to some phase retrieval techniques. We provide here a full approach to wavefront sensing and control. However, we narrow the discussion to the case where aberrations are small, i.e. the problem remains linear. Non-linearity could be dealt with by iterating \cite[(see Give'on, Kasdin, \& Vanderbei 2005)]{Giveon05} or other methods could be used in the first place to get to the wavefront quality needed here.

This paper is organized as follows: in \S\ref{sec:snt} we expose the wavefront sensing and control theory, in \S\ref{sec:sim} we present one- and two-dimensional simulations with white and colored speckle noise, and we conclude in \S\ref{sec:ccl}.

%
%
\section{Speckle nulling theory} \label{sec:snt}
We present here a simplified theory in which diffraction by the pupil is not included. For a more complete exposition of the theory, see \cite{Borde05}.

%
\subsection{Wavefront sensing} \label{sub:wfs}
Let $E(u,v) = e^{i\phi(u,v)} \approx 1 + i\phi(u,v)$ be the normalized electric field due to the star alone in a pupil plane before the coronagraph. Here, $\phi$ is a complex function accounting for small departures from a plane wave, i.e. phase and amplitude aberrations. In order to assess the performance of speckle nulling alone, we assume an ideal coronagraph which perfectly removes the star: in a reimaged pupil after the coronagraph, the field is $E'(u,v) = i\phi(u,v)$. On the detector, the image-plane field is then $\widehat{E'}(\alpha,\beta) = i\widehat{\phi}(\alpha,\beta)$, where hats denote Fourier transforms. Because the detector measures the field intensity, $I(\alpha,\beta) = |\widehat{\phi}(\alpha,\beta)|^2$, complex aberrations cannot be obtained from a single image. Fortunately, in an AO system, the DM provides a natural means to acquire the missing information by modifying the wavefront phase, which leads to a field intensity variation (this is a form of \emph{phase diversity}).

It is easy to show that complex aberrations can be measured from a set of three images, numbered here 0, 1, and 2. Let $\psi$  be the phase change introduced by the DM. Although this is not mandatory, we assume for the sake of simplicity that the DM is originally flat ($\psi_0 = 0$). The intensity takes the successive values
\begin{equation}  \label{eq:syst}
\begin{array}{lcr}
\left \{
\begin{array}{l}
I_0 = |\widehat{\phi}|^2 \\
I_1 = |\widehat{\phi} + \widehat{\psi}_1|^2 \\
I_2 = |\widehat{\phi} + \widehat{\psi}_2|^2 \\
\end{array}
\right.

& \Longleftrightarrow &

\left \{
\begin{array}{l}
{(\widehat{\psi}_1)}^\ast \, (\widehat{\phi}) + (\widehat{\psi}_1) \, {(\widehat{\phi})}^\ast
= I_1 - I_0 - |\widehat{\psi}_1|^2 \\
{(\widehat{\psi}_2)}^\ast \, (\widehat{\phi}) + (\widehat{\psi}_2) \, {(\widehat{\phi})}^\ast
= I_2 - I_0 - |\widehat{\psi}_2|^2. \\
\end{array}
\right. \\
\end{array}
\end{equation}
Provided $\Delta \equiv {(\widehat{\psi}_1)}^\ast \, \widehat{\psi}_2 - \widehat{\psi}_1 \, {(\widehat{\psi}_2)}^\ast \neq 0$, we derive for the complex aberrations
\begin{equation}  \label{eq:phi}
\widehat{\phi} = \frac{\widehat{\psi}_2 \, (I_1 - I_0 - |\widehat{\psi}_1|^2) -
\widehat{\psi}_1 \, (I_2 - I_0 - |\widehat{\psi}_2|^2)}{\Delta}.
\end{equation}
The condition $\Delta \neq 0$ reflects that (\ref{eq:syst}) should not contain twice the same equation, i.e. the DM should be driven so that, between images 1 and 2, the intensity is modified in a measurable way. In practice, $\psi_1$ is computed to modify the intensity in every point by an amount comparable to the intensity in that point. Then, $\psi_2$ is computed to maximize $\Delta$ in every point, so as to maximize the measurement signal-to-noise ratio.

%
%
\subsection{Wavefront control} \label{sub:wfc}
With the knowledge of the complex aberrations, speckles would instantly go away if we could make $\psi$ equal to $-\phi$. However,
\begin{enumerate}
\item the DM cannot take an arbitrary shape as it is made of a facesheet supported by a finite number of actuators \cite[(see Trauger, Moody, Gordon, \etal\ 2003 for a detailed description)]{Trauger03}, and
\item the DM introduces a real, not complex, phase change.
\end{enumerate}
Point (\textit{a}) implies that aberrations can only be corrected up to a maximum spatial frequency set by the density of actuators (with respect to the wavelength), which means that the dark hole cannot be arbitrarily large. Although point (\textit{b}) seems to imply that only phase aberrations could be corrected, it really means -- as we will show later on -- that phase and amplitude aberrations can both be corrected, but at the cost of half the dark hole area.

We model the DM shape as the linear superposition of two-dimensional influence functions. Each of these functions describes the DM surface deformation in response to the actuation of a given actuator. The phase change introduced by a $N\!\times\!N$-element DM is then
\begin{equation}  \label{eq:psi}
\psi(u,v) = \sum_{k=0}^N \sum_{l=0}^N a_{kl} \, f_{kl}(u,v),
\end{equation}
where the $a_{kl}$ (the actuator strokes) are the unknowns in the wavefront control problem.

\subsubsection{Speckle field nulling} \label{subsub:sfn}
As a first approach to the $a_{kl}$ calculation, let us examine under what conditions $\widehat{E'}$ could be made zero in the simplified case where all influence functions have an identical shape, i.e. $f_{kl}(u,v) = f(u-k\frac{d}{\lambda},v-l\frac{d}{\lambda})$, where $d$ is the actuator spacing. We have
\begin{equation}  \label{eq:field}
\widehat{\phi}(\alpha,\beta) + \widehat{\psi}(\alpha,\beta) = 0
\quad \Longleftrightarrow \quad
\sum_{k=0}^{N-1} \sum_{l=0}^{N-1} a_{kl}\,e^{-i \frac{2\pi d}{\lambda} (k \alpha + l \beta)}
= - \frac{\widehat{\phi}(\alpha,\beta)}{\hat{f}(\alpha,\beta)}.
\end{equation}
We recognize in the left-hand size of (\ref{eq:field}) a truncated Fourier series. By setting the strokes equal to the Fourier coefficients of $-\widehat{\phi}/\hat{f}$, we minimize the total energy of $\widehat{E'}/\hat{f}$ in the dark hole. This is close, but not quite the same thing as minimizing the total speckle energy in the dark hole, which would be the total energy of $\widehat{E'}$ (more on that in \S\ref{subsub:sem}). With the Fourier solution to (\ref{eq:field}), we learn that:
\begin{enumerate}
\item because the Fourier series is truncated (as actuators are in limited number), there can be no exact solution to (\ref{eq:field}) and no infinitely deep dark hole, unless the truncated part of the series happen to be equal to zero, i.e. aberrations are band-limited;
\item because Fourier series are periodic, the dark hole will have an extension limited to that period. Precisely, it will be a centered square with a maximum size of $\pm \frac{\lambda}{2d} = \pm \frac{N}{2} \frac{\lambda}{D}$, where $D$ is the pupil's diameter. In other words, a dark hole created with a $N\!\times\!N$-DM will be at most $N\!\times\!N$ element of resolution wide, which is consistent with the sampling theorem.
\end{enumerate}
The solution by Fourier series is attractive because it can be coded with FFTs and is computationally very fast. However, it has two drawbacks: (1) it cannot handle influence function variations across the DM, and (2) it does not exactly minimize the speckle field energy in the dark hole which would seem desirable.

\subsubsection{Speckle energy minimization} \label{subsub:sem}
As a second approach, let us now compute the $a_{kl}$ as the result of the minimization of the total speckle energy in the dark hole, defined as  
\begin{equation}  \label{eq:energy1}
\mathcal{E} \; \equiv \; <\widehat{\phi} + \widehat{\psi}, \widehat{\phi} + \widehat{\psi}>
\; \equiv \; \int\!\!\!\int_\mathcal{H} (\widehat{\phi} + \widehat{\psi}) (\widehat{\phi} + \widehat{\psi})^\ast.
\end{equation}
Given that $\partial \widehat{\psi}/\partial a_{kl} = \hat{f}_{kl}$, the total energy is minimized when, $\forall (k,l)$,
\begin{equation}  \label{eq:energy2}
\frac{\partial \mathcal{E}}{\partial a_{kl}} = 0
\quad \Longleftrightarrow \quad
\sum_{n=0}^{N-1} \sum_{m=0}^{N-1} a_{nm} \, \Re \left(<\hat{f}_{nm},\hat{f}_{kl}> \right) \; = - \Re \left( <\widehat{\phi},\hat{f}_{kl}> \right),
\end{equation}
where $\Re$ stands for the real part.

We recognize in (\ref{eq:energy2}) a linear system that can be solved in matrix format:
\begin{equation}  \label{eq:matrix1}
FA = \Phi \Leftrightarrow A = F^{-1} \Phi,
\end{equation}
where $F$ is a $N^2\!\times\!N^2$-matrix, and $A$ and $\Phi$ are $N^2$-vectors.

It happens that if influence functions have the property of being separable, that is to say if $f_{kl}(u,v) = g_k(u)g_l(v)$, then (\ref{eq:matrix1}) can be rewritten as
\begin{equation}  \label{eq:matrix2}
G \, A \, G = \Phi \Leftrightarrow A = G^{-1} \, \Phi \, G^{-1},
\end{equation}
where $G$, $A$ and $\Phi$ are now $N\!\times\!N$-matrices. This has the advantage of requiring much less computing power. Box-shape functions and bidimensional Gaussians are two examples of fully separable functions. If instrumental influence functions are not fully separable but the wavefront is corrected as if they were, a degradation of the dark hole performance is to be expected (see \S\ref{sub:csn} for an example).

When phase and amplitude aberrations are simultaneously present, $\phi$ is complex, and the careless use of any of the two algorithms would lead to complex values for the $a_{kl}$. The key to correct amplitude aberrations with real actuator strokes is to treat them as phase aberrations. This can be done by deriving from $\widehat{\phi}$ a function with Hermitian symmetry and therefore a real Fourier transform. However, that Hermitian function can be made equal to $\widehat{\phi}$ only in half the dark hole, therefore half the dark hole area is sacrificed.

%
%
\section{Simulations} \label{sec:sim}
%
%
\subsection{White speckle noise} \label{sub:wsn}
\begin{figure}
\centering
 \includegraphics[width=10cm]{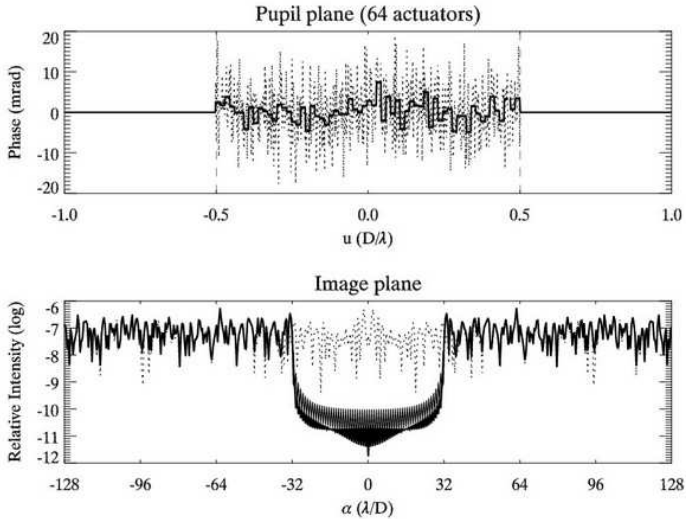}
  \caption{One-dimensional speckle nulling simulation. Top: Phase-aberrated wavefront (dotted line) and negative of DM shape (thick solid line). Bottom: Uncorrected coronagraphic image (dotted line), and corrected coronagraphic image with speckle field nulling (thin solid line) and with speckle energy minimization (thick solid line).}
  \label{fig:1D}
\end{figure}

To start with, we simulate our speckle nulling algorithms in one-dimension with white speckle noise (Fig.~\ref{fig:1D}). The standard deviation of phase aberrations is set to $\lambda/1000$ so that the linear approximation of \S\ref{sub:wfs} would remain valid. Refer to \cite{Borde05} for simulations with amplitude aberrations as well. The DM features 64 actuators and top-hat influence functions. As shown in the bottom panel of Fig.~\ref{fig:1D}, energy minimization yields a dark hole slightly deeper than field nulling: $1.4 \times 10^{-11}$ vs. $5.8 \times 10^{-11}$ (intensities are scaled with respect to the peak of the star Airy function). This is always the case by typically a factor of two. It is possible to obtain deeper dark holes with energy minimization but at the cost of smaller angular extensions. For instance, if one accepts to reduce the dark hole size from 64 to 44 resolution elements, the dark hole floor drops down to $2.7 \times 10^{-15}$.
%

%
%
\subsection{Colored speckle noise} \label{sub:csn}
\begin{figure}
 \includegraphics[width=6cm]{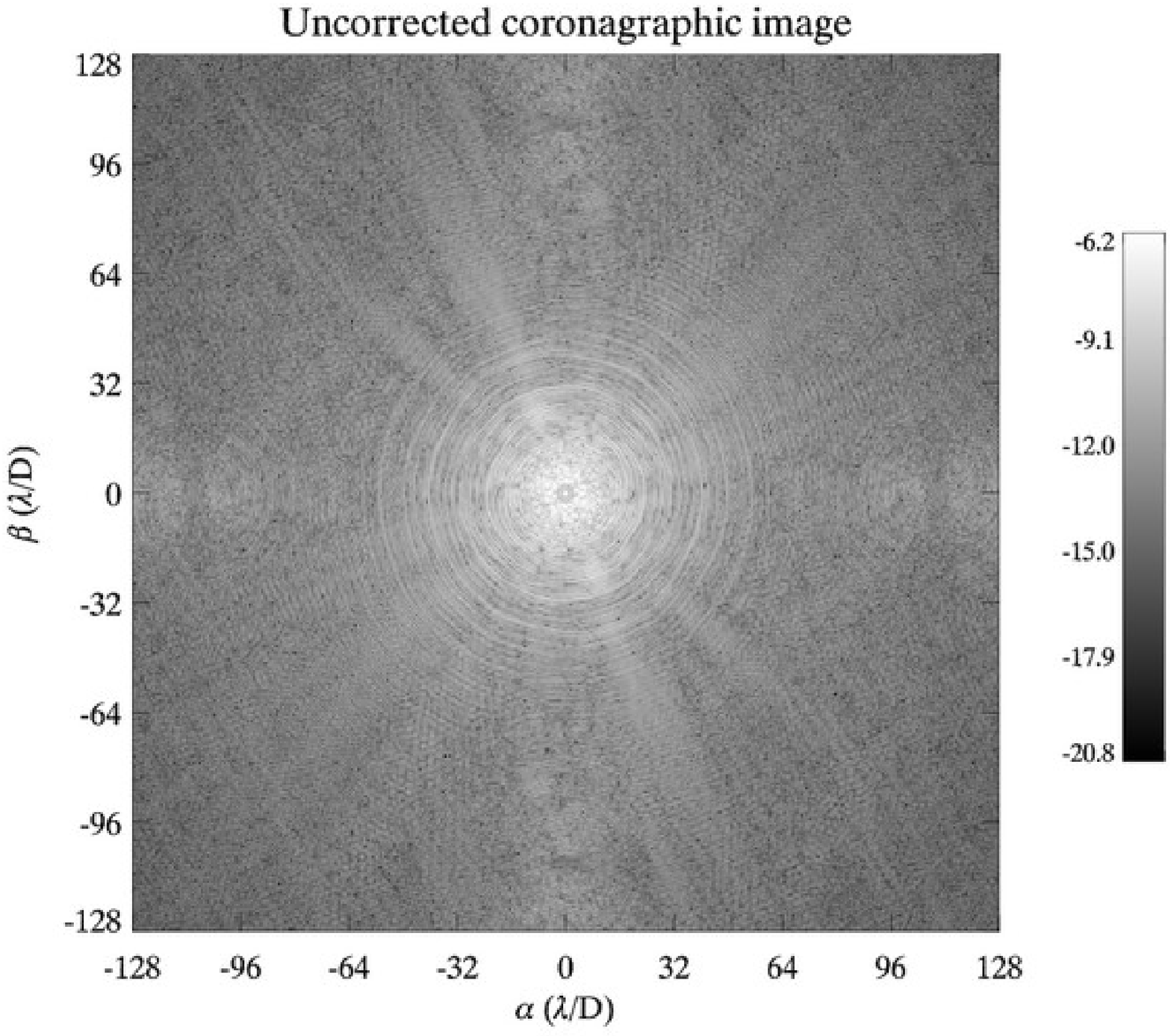}
 \includegraphics[width=6cm]{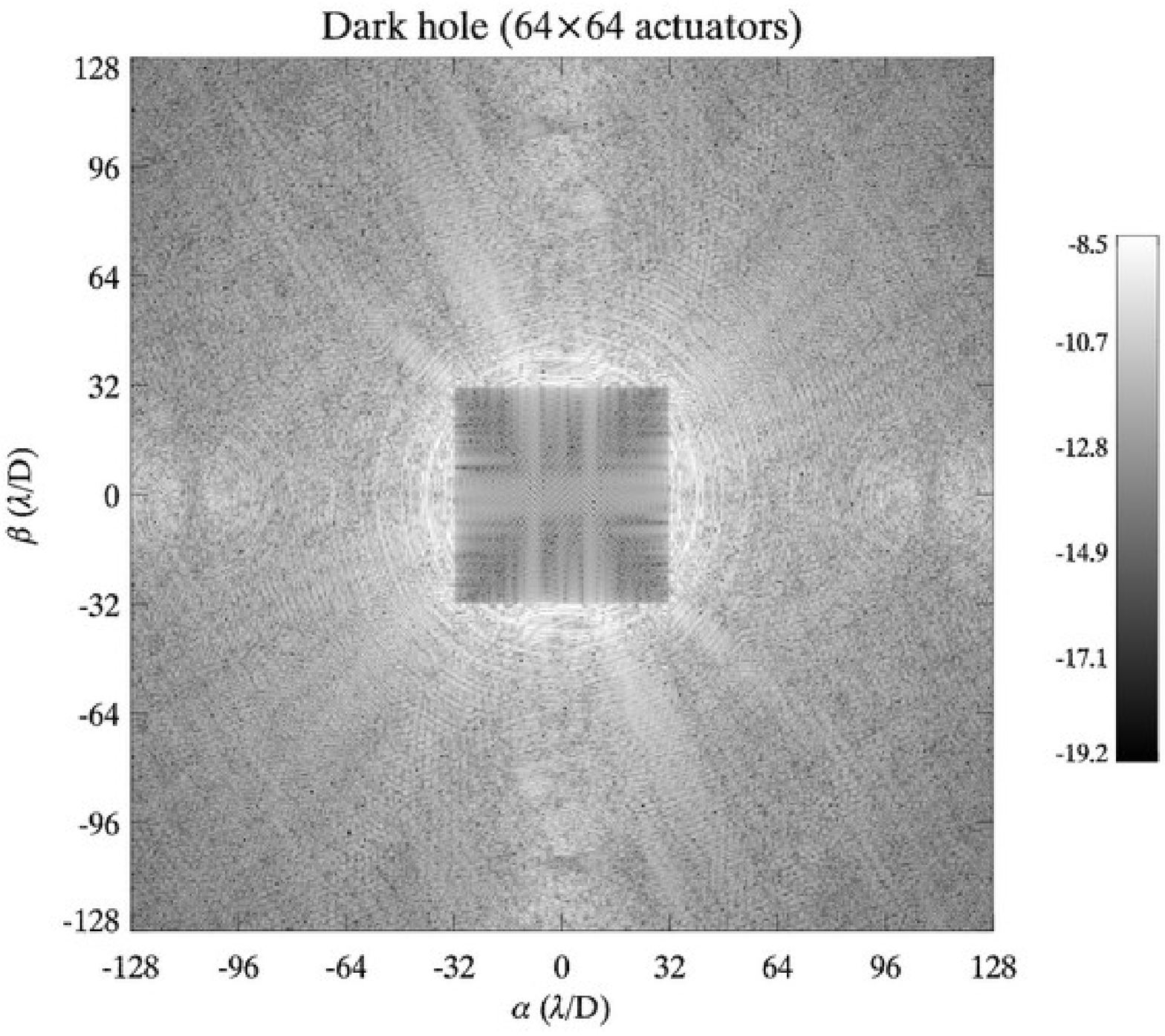}
 \includegraphics[width=6cm]{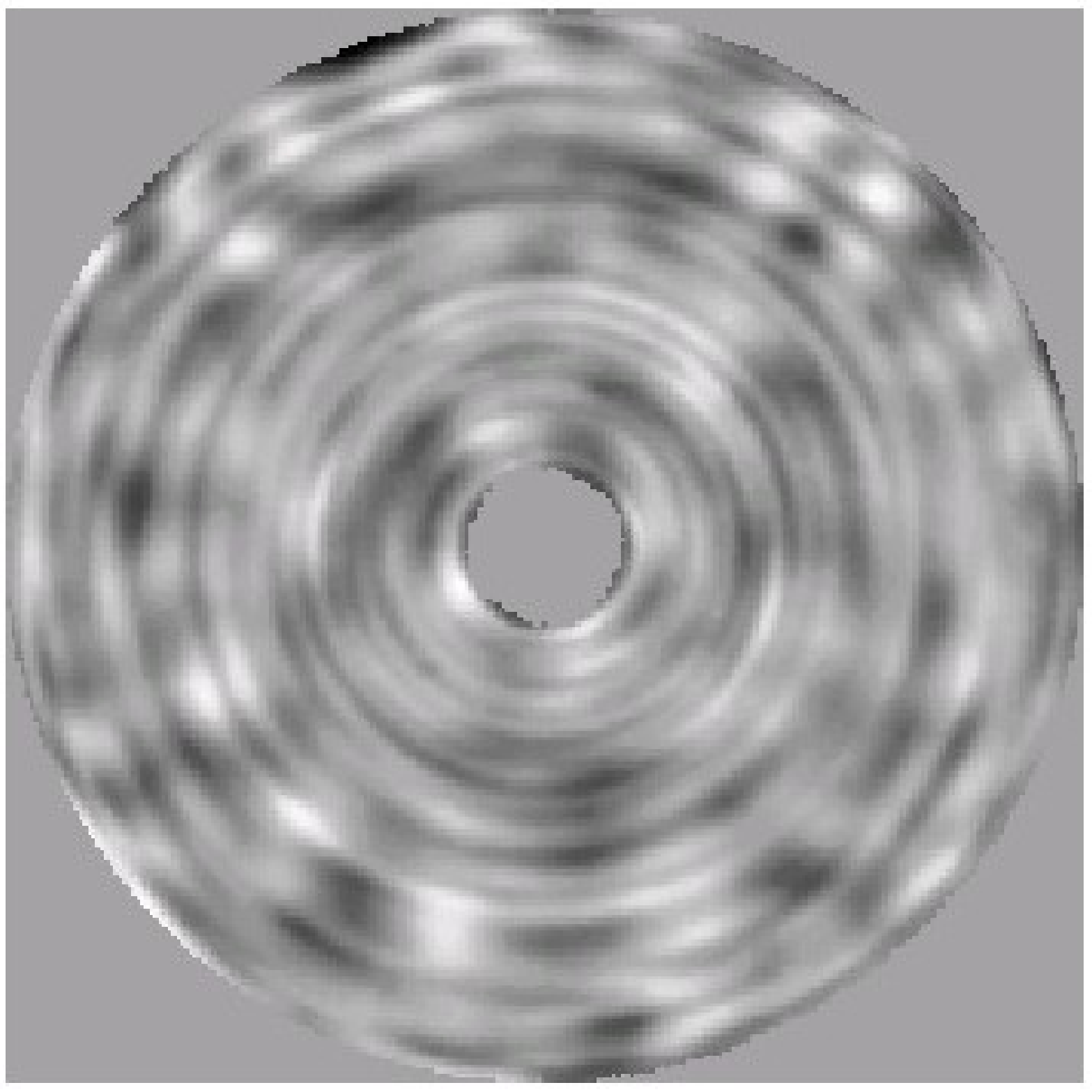} \hspace*{1cm}
 \includegraphics[width=6cm]{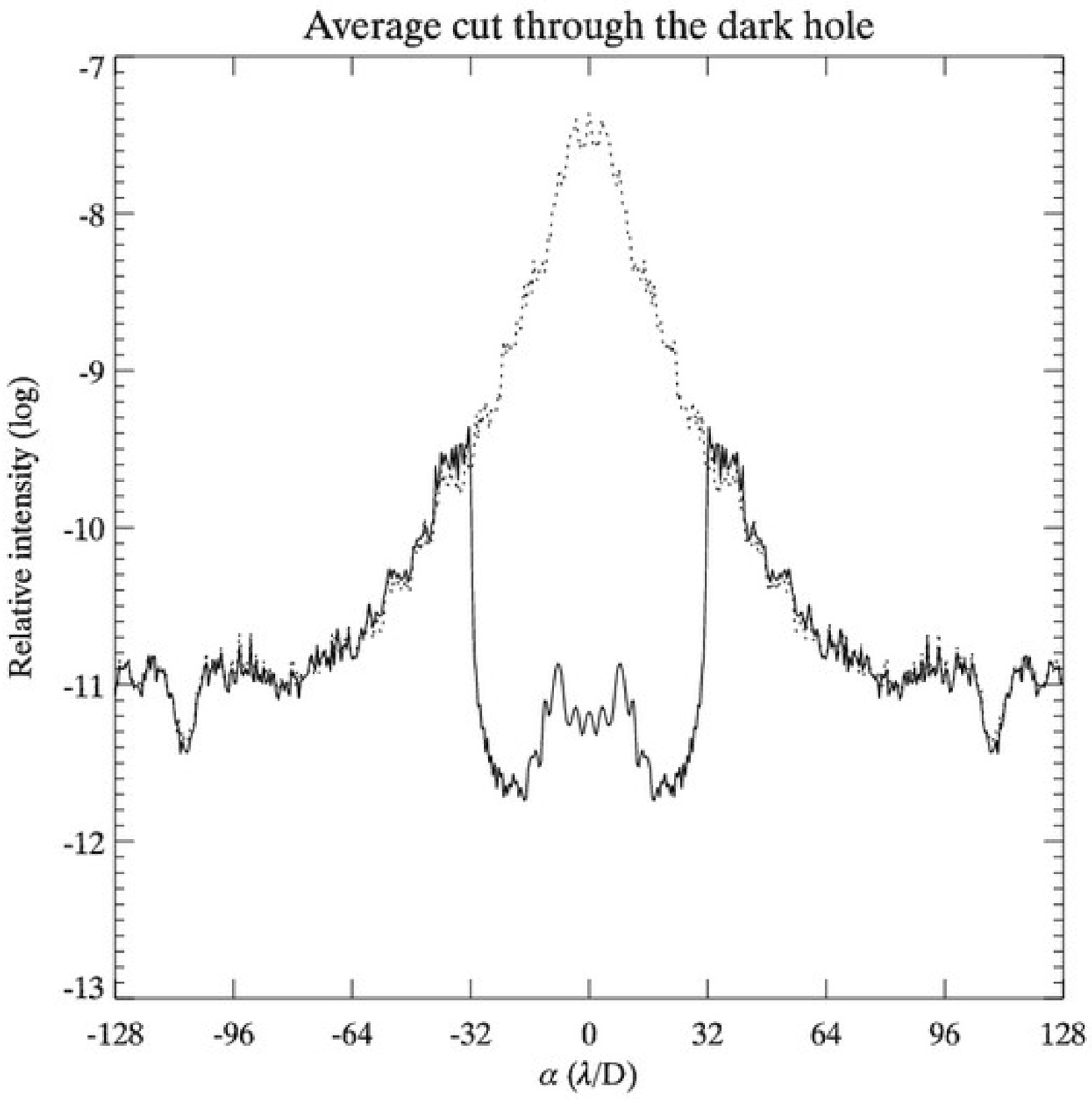}
  \caption{Two-dimensional speckle nulling simulation. Bottom left: phase map of the first 8.2-m primary mirrors of ESO's Very Large Telescope. Top left: uncorrected coronagraphic image obtained with this phase map scaled to a standard deviation of $\lambda/1000$. Top right: corrected image obtained with the field nulling algorithm (\S\ref{subsub:sfn}). Bottom right: average cut through the dark hole.}
  \label{fig:2D}
\end{figure}

Figure~\ref{fig:2D} is the result of a two-dimensional simulation including realistic phase aberrations from the surface map of an 8.2-m mirror. In this map, the power spectral density of surface errors decreases as the third power of the spatial frequency above 0.35~m$^{-1}$. The DM features $64\!\times\!64$ actuators, and influence functions are as measured on the High-Contrast Imaging Testbed \cite[(HCIT; Trauger, Burrows, Gordon, \etal\ 2004)]{Trauger04} at NASA's Jet Propulsion Laboratory. These simulations does not include quantum or read-out noise and assume a perfect DM. With field nulling (as shown in Fig.~\ref{fig:2D}), the average dark hole floor is $5.9 \times 10^{-12}$. Energy minimization assuming separable influence functions yields a higher average floor of $7.1 \times 10^{-11}$, because of a 5\,\% mismatch between the real HCIT functions and their separable approximations. 
%

%
%
\section{Conclusion} \label{sec:ccl}
In this paper, we presented a full theory for wavefront sensing and control using science camera images (no separate wavefront sensing channel), assuming that phase and amplitude aberrations remain small. Sensing is performed with only three images, and relies on the deformable mirror to introduce the needed phase diversity.

Since our theory is general, it can adapted to most coronagraph designs. A good example is the application to shaped pupil coronagraphs by \cite{Giveon05}. We are currently working on the adaptation to coronagraphs with image-plane band-limited masks \cite[(Kuchner \& Traub 2002)]{Kuchner02} to start experimenting with the HCIT. In parallel, we plan to develop a polychromatic theory to accommodate the spectral bandwidth necessary for realistic planet detection and spectroscopy.

%
%
\begin{acknowledgments}
We acknowledge many helpful discussions with Chris Burrows, John Trauger, Joe Green, Stuart Shaklan, Amir Give'on, Anthony Boccaletti, and Pierre Baudoz.  This work was performed in part under contract 1256791 with the Jet Propulsion Laboratory (JPL), funded by NASA through the Michelson Fellowship Program, and in part under contract 1260535 from JPL. JPL is managed for NASA by the California Institute of Technology. This research has made use of NASA's Astrophysics Data System.
\end{acknowledgments}

%
%

%
%
\begin{discussion}

\discuss{Trauger}{Your model assumes that coronagraph optics downstream from the DM do not introduce additional phase and amplitude errors. However, these errors will necessarily exist and affect the image. Could you comment on that?}

\discuss{Bord\'e}{We have indeed assumed an optically perfect coronagraph. I guess that the solution to this problem will be to iterate the algorithm, so that unknown errors will be taken care of in the process. In addition, I should say that the effect of the coronagraph is not only to remove the star PSF as in our simplistic model, but is also to modify the DM influence functions in a way that depends on their positions in the pupil. This can easily be (and will be) incorporated in the energy minimization algorithm, but cannot be dealt with in the field nulling algorithm.}

\end{discussion}

\end{document}